\documentclass{WileyMSP-template}
\usepackage[T1]{fontenc}
\usepackage{amsmath}
\usepackage{cite}
\usepackage{indentfirst}
\usepackage{ragged2e}
\usepackage{color}
\usepackage{bm}

\begin{document}


\title{A large spin-splitting altermagnet designed from the hydroxylated MBene monolayer}

\maketitle

\author{Xinyu Yang}
\author{Shan-Shan Wang*}
\author{Shuai Dong*}

\begin{affiliations}
X. Y. Yang, S.-S. Wang, S. Dong\\
Key Laboratory of Quantum Materials and Devices of Ministry of Education, School of Physics, Southeast University, Nanjing 211189, China\\
E-mail: wangss@seu.edu.cn\\
E-mail: sdong@seu.edu.cn
\end{affiliations}


\justifying
\begin{abstract}
The development of altermagnets is fundamentally important for advancing spintronic device technology, but remains unpractical for the weak spin splitting in most cases, especially in two-dimensional materials. Based on spin group symmetry analysis and first-principles calculations, a novel hydroxyl rotation strategy in collinear antiferromagnets has been proposed to design altermagnets. This approach achieves a large chirality-reversible spin splitting exceeding $1130$ meV in $\alpha_{60}$-Mn$_2$B$_2$(OH)$_2$ monolayer. The system also exhibits intrinsic features of a node-line semimetal in the absence of spin-orbit coupling. Besides, the angles of hydroxyl groups serve as the primary order parameter, which can switch on/off the altermagnetism coupled with the ferroelastic mechanism. The corresponding magnetocrystalline anisotropy have also been modulated. Moreover, an interesting spin-related transport property with the spin-polarized conductivity of 10$^{19}$ $ \Omega^{-1}m^{-1}s^{-1}$ also emerges. These findings uncover the hydroxyl rotation strategy as a versatile tool for designing altermagnetic node-line semimetals and opening new avenues for achieving exotic chemical and physical characteristics associated with large spin splitting.
\end{abstract}

\section{Introduction}

\indent Two-dimensional (2D) magnetism represents a remarkable manifestation of quantum phenomena in condensed matter physics, with researches predominantly focused on two classes of collinear magnetic orders: ferromagnets and antiferromagnets~\cite{gong2017nature,burch2018nature,Xiang2019sci,jungwirth2016NN}. In ferromagnets, the parallel alignment of magnetic moments gives rise to macroscopic magnetization, making them highly suitable for data recording and permanent magnet-based energy storage applications. However, their application in high-density and high-speed information storage is fundamentally constrained by two inherent limitations: unavoidable stray magnetic fields and GHz-speed magnetization dynamics. In contrast, antiferromagnetic materials possess fully compensated magnetic sublattices, enabling superior storage density and ultrafast THz-range spin dynamics that permit picosecond-scale magnetization switching. Nevertheless, the inherent insensitivity to external magnetic fields, coupled with the formidable difficulty in controlling their spin configurations, creates fundamental barriers to device integration and technological implementation~\cite{Jungwirth2018NP,Kamil2018SA,J.2014PRL,P.2016S}.

Altermagnets represent a groundbreaking class of magnetic materials that uniquely combine nonrelativistic spin splitting in the Brillouin zone (BZ) and zero net magnetization in real space~\cite{031042PhysRevX2022,040501PhysRevX2022,BaiAFM2024}. The simultaneous manifestation of these seemingly contradictory properties establishes altermagnets as an available platform for investigating spin-dependent phenomena. Much progress has also been reported towards the three-dimensional (3D) altermagnetic systems, which can give rise to various physical properties, including anomalous Hall effect, large crystal thermal transport and DC Josephson effect~\cite{Zhou2025N,zhou2023PRL,Ouassou2023prl,Krempaský2024N,Song2025NRS}. To gain deeper insights into the fundamental mechanisms of altermagnets and facilitate their integration into spintronic devices, more research efforts have been expanded to investigate two-dimensional (2D) altermagnetic materials. 

Recent theoretical studies have proposed the realization of 2D altermagnets through either direct/reversed stacking or twisted antiferromagnetic bilayer configurations~\cite{sun2025AM,Liuprl2024,sunPRB2024,Sun2024NL}. However, such weak van der Waals (vdW)-interaction-mediated altermagnetic states remain fragile, exhibiting weak band splitting effects due to the inherently weak interlayer coupling. To break free from this fundamental constraint, here we focus on the non-vdW materials. MBenes, a new family of 2D transition metal borides, have a general formula of $M_n$B$_{2n-2}$, where $M$ represents a transition metal, B stands for boron and $n=2$, $3$, $4$. These 2D transition metal borides can be obtained by selectively etching out the A-site layers (Al or In) from their parent $MA$B phases. Based on the crystal symmetries, MBenes can be classified into two distinct types: orthorhombic MBenes and hexagonal MBenes. The corresponding precursors are orthorhombic $MA$B and hexagonal $MA$B, respectively. The MBene family now comprises more than $50$ distinct members, which may provide a fertile zoo to explore 2D altermagnets with relative large spin splittings~\cite{Wang2023acsnano,Nair2022AM,Jakubczak2021AFM,zhang2022jmca}.

In this work, we employ a novel strategy to design an altermagnet in MBene systems via controlling the surface termination rotation. Based on first-principles calculations, we demonstrate that the $\alpha$-Mn$_2$B$_2$(OH)$_2$ monolayer has a polar metal phase with A-type antiferromagnetism. The hydroxyl rotation strategy achieves a transition from the $\alpha$- to $\alpha_{60}$-Mn$_2$B$_2$(OH)$_2$ monolayer, and the $\alpha_{60}$-Mn$_2$B$_2$(OH)$_2$ exhibits altermagnetic characteristics, which also manifests intrinsic features of a node-line semimetal and anisotropic spin-polarized conductivity in the absence of spin-orbit coupling. Consequently, the maximum spin splitting reaches $\sim1130$ meV, superior to most reported 2D altermagnets. Furthermore, our calculations reveal that these two states with spin splitting on/off exhibit strong coupling with the ferroelastic phase transition and magnetocrystalline anisotropy. 
\section{Results and Discussion}

In conventional layered magnetic lattices, in addition to the ferromagnetic (FM) structure, there exist three common antiferromagnetic (AFM) configurations, represented by the intralayer-interlayer magnetic orders: FM-AF (classified as A-type antiferromagnetism, AAFM, with antiferromagnetically coupled FM layers), AF-FM (intralayer AFM coupling and interlayer FM coupling), AF-AF (intralayer and interlayer AFM coupling). 

\begin{figure}[h]
	\includegraphics[width=0.9\linewidth]{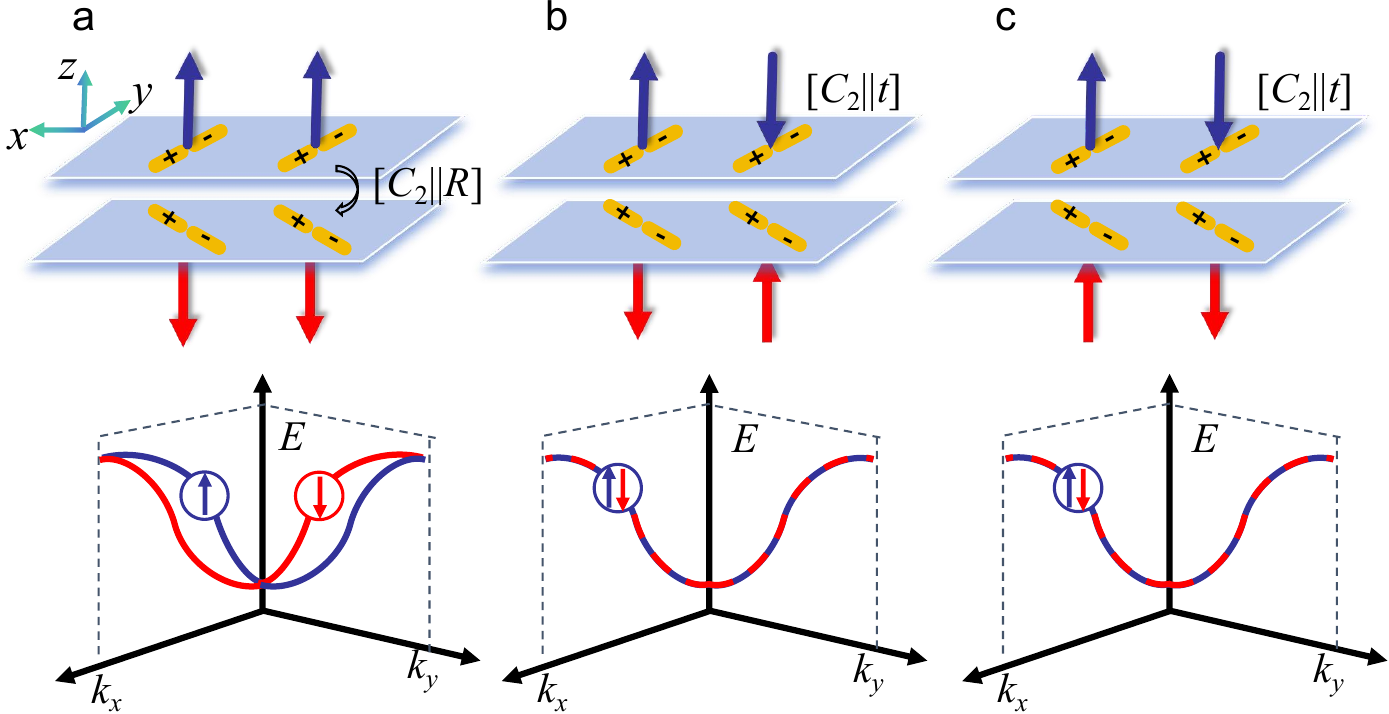}
	\centering
	\caption{Schematic illustration of antiferromagnetic configurations and corresponding band structures. a) FM-AF, classified as AAFM, with the intralayer FM coupling and interlayer AFM coupling. b) AF-AF, with the intralayer and interlayer AFM coupling. c) AF-FM, with the intralayer AFM coupling and interlayer FM coupling. Blue and red arrows: the spins of upper and lower magnetic layers, respectively. Yellow lobes represent the charge polarity, with $+$/$-$ indicating the polar orientation. The transformation $C_2$ refers to the spin-space operation to flip spin up/down, while the transformations ($R$ and $t$) are operations in the real space. $R$: the rotational symmetry operation. $t$: the translational symmetry operation.}
	\label{F1}
\end{figure}

The symmetry analysis based on spin groups serves as a crucial theoretical framework for understanding AFM materials~\cite{031042PhysRevX2022,040501PhysRevX2022}. Spin groups can be mathematically represented as the direct product $r_s \otimes R_s$, where $r_s$ denotes the spin-only group, and $R_s$ represents nontrivial spin groups containing paired transformations [$R_i || R_j$]~\cite{Zeng2024prb}. The left operation $R_i$ purely operates in the spin space. The right operation $R_j$ solely acts in the real space. In the context of layered magnetic lattices, the spin space and real space are decoupled, allowing the AFM configurations to be described using the nontrivial spin groups.

When any lattice distortion breaks the inversion symmetry between vertically stacked magnetic layers, the upper and lower electron clouds will also be distorted correspondingly. Based on the nontrivial spin group theory, the A-type antiferromagnetic configuration will exhibit the altermagnetic characteristics, as shown in \textbf{Figure~\ref{F1}}a. [$C_2||R$] represents the case in which the spin up and down sublattices can be connected by the rotation symmetry $R$. In this schematic diagram, R represents $C_{2x}$, a two-fold rotation around the $x$ axis. This spin group symmetry enables the emergence of alternating spin polarization in momentum space. While the AF-FM and AF-AF configurations maintain conventional antiferromagnetic properties due to that the spin up and down sublattices can be connected by a direct translation symmetry operation $t$, as shown in Figure~\ref{F1}b,c.

In principle, this design strategy generally works for realizing altermagnetism in layered magnets. However, to enhance the spin-splitting, it is better to select those magnets with strong interlayer couplings, which can be realized in some non-vdW layered antiferromagnets. Recent experimental advances have demonstrated the successful synthesis of 2D orthorhombic MnB nanosheets through thermal etching of Mn$_2$AlB$_2$ precursors, and hydroxyl groups (OH) were also introduced after exfoliation~\cite{Wang2023acsnano}. Our first-principles calculations reveal that the hexagonal structural phase is even more stable than the orthorhombic phase by $4.21$ meV/f.u. lower in energy, as compared in Figure~S1 in Supporting Information (SI). The hexagonal MnB monolayer possesses a trigonal lattice with the space group $P6/mmm$, in which B ions are sandwiched between two Mn layers.

After the hydroxyl adsorption, there are two distinct configurations ($g$- and $\alpha$-Mn$_2$B$_2$(OH)$_2$ monolayer)~\cite{Wang2023acsnano,Liu2024PRB}. For the intralayer triangular lattices, there are three common types of magnetic structures: FM, Zigzag-type antiferromagnetism (ZAF) and stripy-type antiferromagnetism (SAF), coupled with the interlayer magnetic orders (FM {\it vs} AFM). There exists six possible magnetic structures (AAFM, FM, SAF-AF, SAF-FM, ZAF-AF, and ZAF-FM), as shown in Figures~S2 and S3 of SI. According to our calculation, the AAFM order of $g$-Mn$_2$B$_2$(OH)$_2$ has the lowest energy, as shown in \textbf{Table~1}. Even though, the $\alpha$-Mn$_2$B$_2$(OH)$_2$ can also stabilize in the AAFM state, whose energy is merely $34$ meV/f.u. higher than the $g$-Mn$_2$B$_2$(OH)$_2$. And the phonon spectrum calculations confirm the dynamical stability for both the $g$- and $\alpha$-Mn$_2$B$_2$(OH)$_2$, with no apparent imaginary vibration modes (Figure~S4 in SI). In addition, the energy barrier between these two states was relatively high ($\sim1.43$ eV/f.u.), indicating that both of them would remain stable within their own potential well regions, see Figure~S5 in SI.

\begin{table}
	 \centering
	\caption{Calculated energies (in units of meV/f.u.) and space group (S.G.) of different models ($g$- and $\alpha$-Mn$_2$B$_2$(OH)$_2$) with six possible magnetic orders. The $g$-Mn$_2$B$_2$(OH)$_2$ with AAFM order is taken as the reference, which owns the lowest energy.}
	\label{Table 1}
	\begin{tabular}[htbp]{ccccc}
		\hline
		Order & Energy ($g$-state)&S.G & Energy ($\alpha$-state)& S.G.\\
		\hline
		AAFM    & 0 &$P\bar{3}m1$ & 34& $Amm2$   \\		
		FM      &58 &$P\bar{3}m1$&222 &$Amm2$  \\
		SAF-AF  &34 &$C2/m$&51 &$Amm2$\\
		SAF-FM  &53&$C2/m$&101 &$Amm2$ \\
		ZAF-AF  &41 &$P2/c$   & 95& $Pmc21$\\
		ZAF-FM  &49& $P2/c$   & 123& $Pmc21$\\
		\hline
	\end{tabular}
\end{table}

\begin{figure}
	\includegraphics[width=\linewidth]{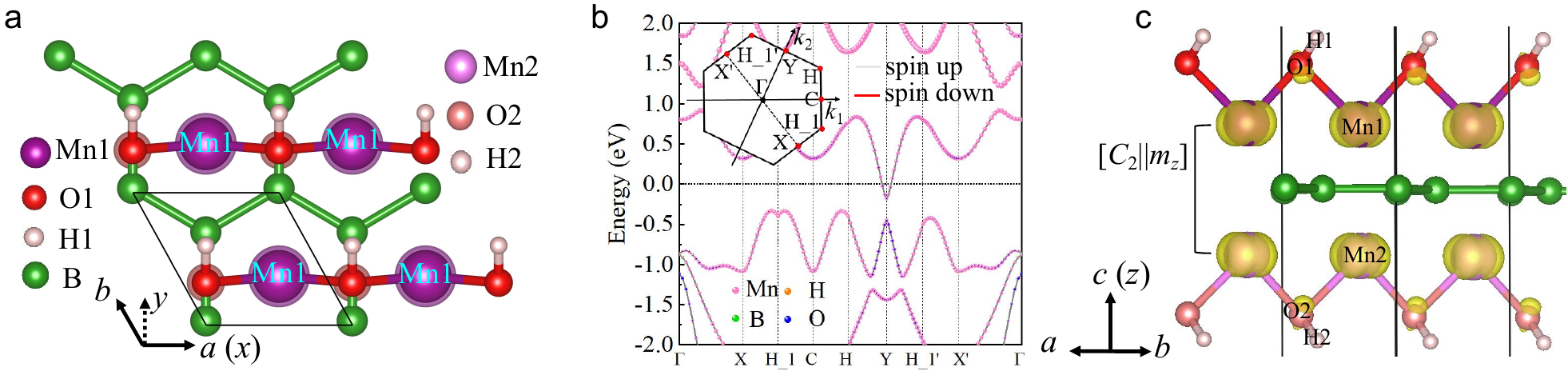}
	\caption{Properties of $\alpha$-Mn$_2$B$_2$(OH)$_2$ monolayer. a) The top view. The primitive cell is indicated by the black-line rhombus. b) Electronic band structure without spin splitting. Inset: the Brillouin zone and high symmetry points. c) Side view of valence electron distribution, integrated within [$-0.5$, $0$] eV.}
	\label{F2}
\end{figure}

Following analysis will focus on the $\alpha$-Mn$_2$B$_2$(OH)$_2$ monolayer with broken space inversion symmetry, which exhibits a space group $Amm2$. There are six equivalent states for the $\alpha$-Mn$_2$B$_2$(OH)$_2$ due to different orientations of hydroxyl groups (see Figure~S6 in SI). As shown in \textbf{Figure~\ref{F2}}a, the calculated lattice parameters ($a=3.038$ \AA{} and $b=3.135$ \AA{}) with an interaxial angle of $\gamma=118.98^\circ$ show agreement with previously reported values~\cite{Liu2024PRB}. The magnetocrystalline anisotropy is further investigated by rotating the spin orientation. As illustrated in Figure~S7 in SI, the magnetic easy axis is oriented along the $y$ axis. The calculated magnetocrystalline anisotropy energy (MAE) reaches approximately $100$ $\mu$eV per Mn, indicating a weak anisotropy.

Based on the space group symmetry analysis, this primitive cell preserves mirror symmetries $m_x$, $m_z$ and a two-fold rotation symmetry $C_{2y}$, which enforce a net polarization along the $y$ direction originating from the hydroxyl radical displacements. Our first-principles calculations reveal that the conduction band minimum (CBM) is primarily contributed by Mn's $3d$ orbitals, intersects with the Fermi level, as illustrated in Figure~\ref{F2}b. This rare coexistence of polarity and metallicity may offer significant multifunctional applications~\cite{Kim2016nature,Puggioni2014nc,Shi2013nm}. The band structure maintains complete spin degeneracy across the BZ path $\Gamma$-X-H$\_$1-C-H-Y-H$\_1'$-X$'$-$\Gamma$. This spin degeneracy is protected by the operations of nontrivial spin group symmetry [$C_2||m_z$], in which the magnetic sublattices can be connected by the mirror symmetry through the $xy$ plane, as shown in Figure ~\ref{F2}c~\cite{Zeng2024prb}.

\begin{figure} [h]	\includegraphics[width=\linewidth]{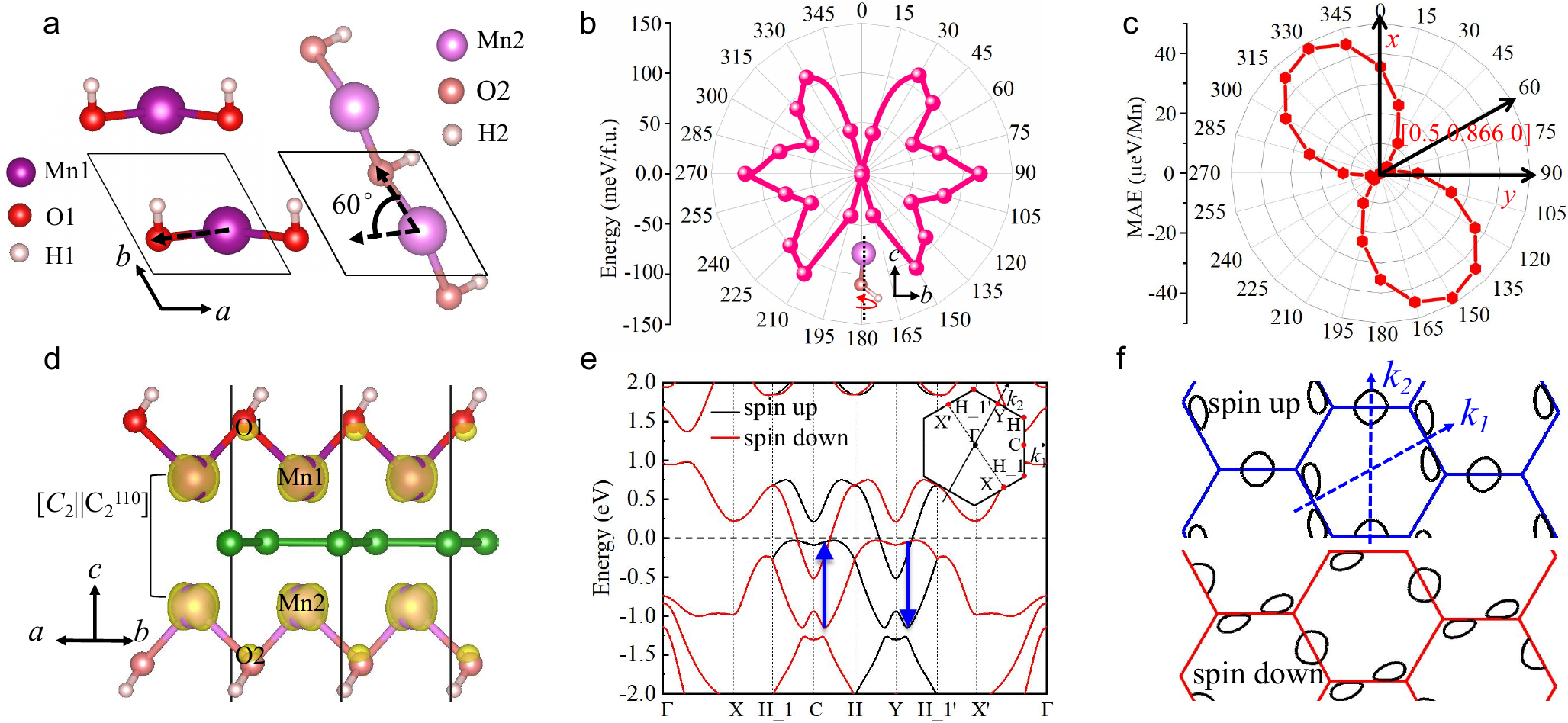}
	\centering
	\caption{a) Top view of rotation of one side hydroxyl groups, which leads to the $\alpha_{60}$-Mn$_2$B$_2$(OH)$_2$ monolayer. Left: the upper layer. Right: the lower layer. The primitive cell is indicated by the black-line rhombus. b) The energy evolution during the in-plane rotation of one layer OH around the $c$ axis, characterized by the rotation angle $\varphi$. When $\varphi=180^{\circ}$, it reaches the $\alpha_{180}$-Mn$_2$B$_2$(OH)$_2$, namely the orientations of hydroxyl groups on the upper and lower surfaces are completely opposite. The energy of $\alpha_{180}$-Mn$_2$B$_2$(OH)$_2$ is very close to (only $0.1$ meV/u.c. higher) that of $\alpha$-Mn$_2$B$_2$(OH)$_2$. c) The in-plane MAE as a function of spin orientation in the $\alpha_{60}$-Mn$_2$B$_2$(OH)$_2$ monolayer. d) Side view of electron density near the Fermi level in the $\alpha_{60}$-Mn$_2$B$_2$(OH)$_2$ monolayer, integrated within [$-0.5$, $0$] eV. The [110] direction corresponds to the diagonal between the $a$-axis and $b$-axis. e) The spin-resolved band structure without SOC of the $\alpha_{60}$-Mn$_2$B$_2$(OH)$_2$ monolayer. f) The spin-up and spin-down Fermi surfaces.}
	\label{F3}
\end{figure}

Following the aforementioned design principle, here we propose a previously unreported strategy for achieving altermagnetism. As illustrated in \textbf{Figure~\ref{F3}}a, when the hydroxyl groups at one side rotate around the $c$ axis, a twisting angle will appear between the upper and lower Mn-O bonds projected onto the $ab$ plane. The energy evolution as a function of twisting angle $\varphi$ is shown in Figure~\ref{F3}b, implying metastable states at $\sim\pm60^\circ$, $\sim\pm120^\circ$, and $180^\circ$, named as $\alpha_{60}$-,  $\alpha_{120}$- and $\alpha_{180}$-Mn$_2$B$_2$(OH)$_2$, respectively.

For the $\alpha_{60}$-Mn$_2$B$_2$(OH)$_2$ monolayer, AAFM remains the magnetic ground state after the full structural optimization, as shown in Table~S1 in SI. Based on the optimized ground-state structure, the exchange coupling parameters are calculated by mapping the DFT energies to the Heisenberg model~\cite{ZhaiAFM2024,yang2024prb}, the exchanges interactions coefficients of the nearest-neighbor $J_1$, next-nearest-neighbor $J_2$, and next-next-nearest-neighbor $J_3$ are estimated as -11.1, 63.3, and -11.9 meV, respectively. The intralayer 1$^{st}/2^{nd}$ neighboring exchanges ($J_1$, $J_3$) guarantee the intralayer collinear ferromagnetic order. The interlayer exchange ($J_2$) prefers layered antiferromagnetism. Therefore, this combination of interactions are not frustrated, which collectively stabilize the AAFM order, as shown in Figure~S8 of SI. The phonon spectrum also maintains its dynamical stability, as evidenced by no apparent imaginary frequencies (Figure~S9 in SI).  However, the MAE changes a lot, with an easy plane [$\sqrt{3}$, $-1$, $0$], as shown in Figures~\ref{F3}c and S10 in SI. It has $C2$ space group and $C2.1$ magnetic space group. The electron distribution near the Fermi level is also distorted between the upper and lower layers, as shown in  Figure~\ref{F3}d. Now the upper and lower Mn sites with opposite spins are connected by the [$C_2||C_2^{110}$] symmetry. This symmetry condition allows momentum-space alternating spin polarization, despite the complete spin-up/spin-down cancellation in the density of states (Figure~S11 in SI). 

The electronic band structure of $\alpha_{60}$-Mn$_2$B$_2$(OH)$_2$ indeed demonstrates the existence of spin splitting without spin-orbit coupling (SOC), as shown in Figure~\ref{F3}e. The bands remain spin-degenerated along the paths $\Gamma$-X-H$\_$1 and H$\_1'$-X$'$-$\Gamma$. However, once deviating from these high-symmetry paths, the spin splitting emerges for those bands close to the Fermi level. Moreover, the spin-up and spin-down Fermi surfaces in the $k_z=0$ plane are also related by the rotational symmetry, exhibiting the band alternating phenomena, as shown in Figure~\ref{F3}f.

\begin{figure}[h]
	\includegraphics[width=0.85\linewidth]{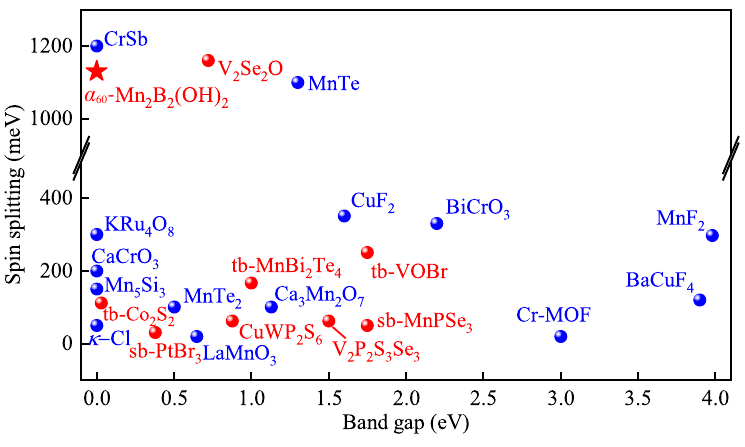}
	\centering
	\caption{The values of maximum spin splitting and band gaps for reported altermagnetic candidates. 2D and 3D materials are labeled in red and blue, respectively. tb is the abbreviation for twisted bilayer. sb stands for the stacking bilayer.}
	\label{F4}
\end{figure}

The alternating large band splitting regions are highlighted with blue arrows in Figure~\ref{F3}e. The maximum splitting reaches $1130$ meV, which surpasses most reported materials, as compared in \textbf{Figure~\ref{F4}}~\cite{sun2025multifield,Duan2025PRL,Liuprl2024,sunPRB2024,040501PhysRevX2022,Gu2025prl,Ma2021nc,Zhu2024nature,Garcia2018prl,sun2025AM}. Although the metallic CrSb shows an even slightly larger band splitting~\cite{031042PhysRevX2022,Zhou2025N}, it is a 3D bulk while 2D monolayer owns the inborn superiority for integration. Following this design strategy, we have also provide additional altermagnetic examples $\alpha_{60}$-Cr$_2$B$_2$(OH)$_2$ and $\alpha_{60}$-Fe$_2$B$_2$(OH)$_2$, as shown in Figure~S12 of SI.

Additionally, the $\alpha_{60}$-Mn$_2$B$_2$(OH)$_2$ also exhibits nodal-line semimetal features without SOC. As shown in Figure~\ref{F3}e, spin-polarized band crossings occur between the valence and conduction bands near the Fermi level, around the high-symmetry points C (0.5, 0, 0) and Y (0 0.5 0). The corresponding 3D plot demonstrates the band crossings form nodal lines, as shown in Figure~S13 of SI.

\begin{figure}
	\includegraphics[width=\linewidth]{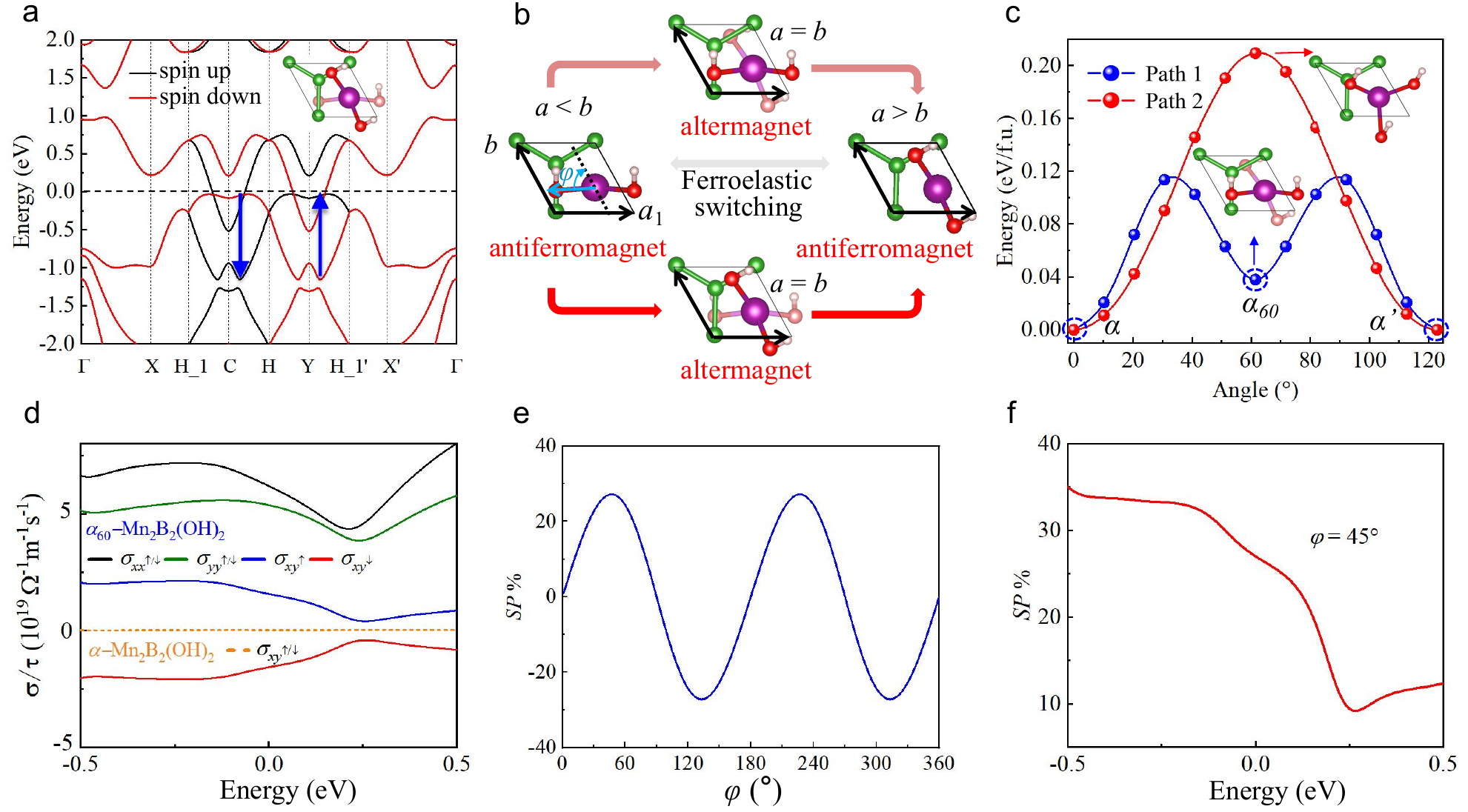}
	\centering
	\caption{a) The spin-resolved band structure without SOC in the $\alpha_{60}'$-Mn$_2$B$_2$(OH)$_2$ monolayer. b) Schematic diagram of the ferroelastic transition of Mn$_2$B$_2$(OH)$_2$ monolayer, along with the transformation of the magnetic properties. c) Energy barriers for two possible paths for the ferroelastic switching. The horizontal axis is defined by the sum of two angles of upper and lower hydroxyl groups ($\varphi_1$ and $\varphi_2$). Path I: rotate one layer of hydroxyl groups first ($\varphi_1$=$60^{\circ}$ and $\varphi_2$=$0^{\circ}$), then the other layer ($\varphi_1$=$60^{\circ}$ and $\varphi_2$=$60^{\circ}$). Path II: simultaneously rotate both the upper and lower hydroxyl groups ($\varphi_1$=$60^{\circ}$ and $\varphi_2$=$60^{\circ}$). {d) Spin-resolved conductivity of the altermagnetic $\alpha_{60}$-Mn$_2$B$_2$(OH)$_2$ monolayer. For comparison, the spin-resolved conductivity of the conventional collinear antiferromagnetic $\alpha$-Mn$_2$B$_2$(OH)$_2$ monolayer has also been calculated, denoted as the orange dashed line. e) Spin polarization $SP$ at 0 eV as a function of $\varphi$. Here, $SP$= $\sigma_{xy}^\uparrow$sin(2$\varphi$)/($\sigma_{xx}^\uparrow$cos$^2$$\varphi$+$\sigma_{yy}^\uparrow$sin$^2$$\varphi$). f) Spin polarization $SP$ at $\varphi$=45$^\circ$ as a function of energy.}}
	\label{F5}
\end{figure}

Upon a $60^{\circ}$ rotation of the upper hydroxyl groups in $\alpha$-Mn$_2$B$_2$(OH)$_2$, the altermagnetic node-line semimetal $\alpha_{60}'$-Mn$_2$B$_2$(OH)$_2$ can be obtained, which exhibits an inverted spin splitting compared to $\alpha_{60}$-Mn$_2$B$_2$(OH)$_2$, as shown in \textbf{Figure~\ref{F5}}a. 
In addition, the aforementioned $\alpha_{120}$-Mn$_2$B$_2$(OH)$_2$ with AAFM order also exhibits the alternating spin-splitting band structures, with the maximum splitting $1110$ meV, as shown in Figure~S14 in SI. Regarding other antiferromagnetic configurations, such as SAF-AF, it preserves the spin degeneracy protected by [$C_2||t$] symmetry (Figure~S15 in SI), also included in our design principle (Figure~\ref{F1}). 

Ferroelasticity is defined by the existence of several equally stable orientation variants that allow the change from one variant to another~\cite{Wu2022nc,Liu2023prb}. As mentioned before, there are six equivalent $\alpha$-Mn$_2$B$_2$(OH)$_2$. We examine two representative cases, $\alpha$- and $\alpha'$-Mn$_2$B$_2$(OH)$_2$. These two equivalent ground states have degenerated energy and interchanged lattice constants of $a$ and $b$. Figure~\ref{F5}b shows the initial state ($\alpha$) and final state ($\alpha'$). The ferroelastic transition process may go through the  $\alpha_{60}$- or $\alpha_{60}'$-Mn$_2$B$_2$(OH)$_2$ monolayer (i.e. the path I), whose lattice constants are $a=b=3.084$ \AA, an approximate average value of the initial $a$ and $b$. Here the angles of hydroxyl groups serve as the primary order parameter, which can switch on/off the altermagnetism coupled with the ferroelastic mechanism. However, there are also other possible ferroelastic switching paths. For example, the upper and lower hydroxyl groups may rotate simultaneously (i.e. the path II). As shown in Figure~\ref{F5}c, the energy barrier of path I is significantly lower than that of that of path II. Therefore, the altermagnetic $\alpha_{60}$-Mn$_2$B$_2$(OH)$_2$ seems to be highly possible during the ferroelastic switching of Mn$_2$B$_2$(OH)$_2$ monolayer.

Another interesting spin-related transport property, i.e. the anisotropic spin-polarized conductivity without spin-orbit coupling, emerges in altermagnets, which is absent in conventional collinear antiferromagnets. The  $\alpha_{60}$-Mn$_2$B$_2$(OH)$_2$ monolayer possesses spin point group $^22^2m$, and according to the Boltzmann transport theory, its symmetry element  [$C_2||C_2^{110}$] gives rise to the relationship of spin conductivity tensors: $\sigma_{xy}^\uparrow=-\sigma_{xy}^\downarrow$, $\sigma_{xx}^\uparrow=\sigma_{xx}^\downarrow$, and $\sigma_{yy}^\uparrow=\sigma_{yy}^\downarrow$ ($x$, $y$, $z$ for Cartesian components and $\uparrow$, $\downarrow$ for spin index)~\cite{PRL2021Gonz,Ma2021nc,Dou2025prb}. Figure 5d shows our calculated spin-resolved conductivity as a function of energy respect to Fermi level. It is noted that the diagonal components $\sigma_{xx}$ and $\sigma_{yy}$ are spin-degenerated while the nondiagonal component $\sigma_{xy}$ is spin polarized, which are consistent with the spin point group symmetry. The magnitude of spin-resolved conductivity has reach to $10^{19}\Omega^{-1}m^{-1}s^{-1}$, which is comparable with that of Mn$_4$(PO$_4$)$_3$ ($\sim$9$\times$$10^{18}$ $\Omega^{-1}m^{-1}s^{-1}$) and RuO$_2$ ($\sim$5.7$\times$10$^{19}$ $\Omega^{-1}m^{-1}s^{-1}$)~\cite{Yang2025AFM,PRL2021Gonz}. However, this spin-polarized conductivity $\sigma_{xy}^s$ is absent in the conventional collinear antiferromagnet ($\alpha$-Mn$_2$B$_2$(OH)$_2$), as indicated by the orange dashed line in Figure 5e.

As for the $\alpha_{60}$-Mn$_2$B$_2$(OH)$_2$ monolayer, the spin resolved conductivity $\sigma_{\bm{nn}}^s$ along the direction $\bm{n}$=(cos$\varphi$, sin$\varphi$, 0) is calculated as $\sigma_{\bm{nn}}^s$=$\sigma_{xx}^s$cos$^2$$\varphi$ + $\sigma_{yy}^s$sin$^2$$\varphi$+ $\sigma_{xy}^s$sin2$\varphi$, where $s$ denotes spin index $\uparrow$ and $\downarrow$, $\varphi$ denotes the azimuthal angle (range: [0, 2$\pi$]). The anisotropic spin polarization $SP$ is defined as $SP$ =($\sigma_{\bm{nn}}^\uparrow$-$\sigma_{\bm{nn}}^\downarrow$)/(($\sigma_{\bm{nn}}^\uparrow$+$\sigma_{\bm{nn}}^\downarrow$). Accordingly, the $SP$ is calculated, as shown in Figure 5e, exhibiting the significant spatial anisotropy. The maximum magnitude of $SP$ reaches nearly 30$\%$ at $\varphi$=45$^\circ$ around the Fermi level, as shown in Figure 5f.

\section{Conclusion}
Based on first-principles calculations, we have uncovered a unique strategy for designing altermagnets through surface termination rotation in the non-vdW MBene systems. Our results demonstrate that the $\alpha_{60}$-Mn$_2$B$_2$(OH)$_2$ with intrinsic node-line semimetal property can be obtained from rotating the hydroxyl groups in $\alpha$-Mn$_2$B$_2$(OH)$_2$ monolayer, which exhibits the alternating spin polarization achieving 1130 meV spin splitting. The predicted giant splitting exceeds the most previously reported altermagnets, suggesting a promising platform for spintronic devices. Besides, this rotational mechanism is closely associated with a ferroelastic phase transition, which simultaneously controls the spin-splitting on/off states and magnetocrystalline anisotropy. Furthermore, a significant spin polarization has also been investigated in the altermagnetic $\alpha_{60}$-Mn$_2$B$_2$(OH)$_2$ monolayer. These findings demonstrate that our research opens a promising avenue for future studies of altermagnetic properties, offering enhanced performance and scalability for advanced spintronic applications.

\section{Experimental Section}
First-principles calculations based on density functional theory (DFT) are performed with the projector augmented-wave (PAW) pseudopotentials as implemented in the Vienna {\it ab initio} Simulation Package (VASP) ~\cite{kresse1996Prb}. The exchange-correlation functional is treated using Perdew-Burke-Ernzerhof (PBE) parametrization of the generalized gradient approximation (GGA)~\cite{perdew1996Prl}. A vacuum space of $20$ \AA{} thickness is added along the $c$-axis direction to avoid layer interactions. The energy cutoff is fixed to $520$ eV. The $\Gamma$-centered $11\times11\times1$ Monkhorst-Pack \textit{k}-mesh is adopted. The convergence criterion for the energy is $10^{-7}$ eV for self-consistent iteration, and the Hellman-Feynman force is set to $0.001$ eV/\AA{} during the structural optimization. As reported previously, the Hubbard correction is considered using the GGA+$U$ method introduced by Liechtenstein {\it et al.}, with $U=3$ eV and $J=1$ eV imposed on Mn's $3d$ orbitals~\cite{Liu2024PRB}. Phonopy is adopted to calculate the phonon band structures~\cite{TOGO20151SCR}. The Fermi surface is plotted using the FermiSurfer program \cite{KAWAMURA2019197cpc}. The spin-resolved conductivity is computed within the framework of Boltzmann transport theory using the relaxation time $\tau$ approximation, as implemented in the BoltzWann module~\cite{PIZZI2014CPC} of the WANNIER90 package ~\cite{Pizzi_2020JPCM,MOSTOFI2008Arash}. For the conductivity calculations, we employ a temperature of 300 K in the Fermi distribution function, and the $k$-point mesh 200$\times$300$\times$1 is used for the conventional cells of $\alpha_{60}$-Mn$_2$B$_2$(OH)$_2$ monolayer ($a$=5.306 \AA{} and $b$=3.145 \AA{}) and $\alpha$-Mn$_2$B$_2$(OH)$_2$ monolayer ($a$=5.486 \AA{} and $b$=3.038 \AA{}).

\medskip
\textbf{Supporting Information}

Supporting Information is available from the Wiley Online Library or from the author.

\medskip
\textbf{Acknowledgements} \par 
	This work was supported by National Natural Science Foundation of China (Grant Nos. 12325401 \& 12274069), the Fundamental Research Funds for the Central Universities (Grant No. 2242025K30023), the Postgraduate Research \& Practice Innovation Program of Jiangsu Province (Grant No. KYCX24\_0361), acknowledges the support of the open research fund of Key Laboratory of Quantum Materials and Devices (Southeast University). Most calculations were done on the Big Data Computing Center of Southeast University.

\bibliographystyle{elsarticle-num}
\bibliography{reference.bib}

\end{document}